\documentclass[aps,twocolumn,pra,superscriptaddress]{revtex4-2}
\usepackage{amsmath,amssymb,amsfonts,bbm,graphicx,times,psfrag}
\usepackage[pdftex]{color}
\usepackage{lineno}

\usepackage{amsmath,bm}
\usepackage[colorlinks, linkcolor=blue, citecolor=blue, urlcolor=blue, breaklinks=true]{hyperref}
\usepackage{microtype}

\usepackage{amsthm}
\usepackage{csquotes}
\usepackage{amsthm}\newcommand{\be} {\begin{equation}}
\newcommand{\ee} {\end{equation}}

\usepackage{color}
\usepackage{mathrsfs}

\newcommand{\dd}{{\rm d}}

\newcommand{\vc}{{\bf c}}

\newcommand{\ic}{{i}}
\newcommand{\e}{{e}}

\begin{document}

\title{Achieving Strong Magnon Blockade through Magnon Squeezing in a Cavity Magnetomechanical System}
\author{M. Amazioug}
\affiliation{LPTHE, Department of Physics, Faculty of sciences, Ibnou Zohr University, Agadir, Morocco.}
\author{D. S. Dutykh}
\affiliation{Department of Mathematics, Khalifa University, Abu Dhabi 127788, United Arab Emirates.}
\author{B. Teklu}
\affiliation{Department of Mathematics, Khalifa University, Abu Dhabi 127788, United Arab Emirates.}
\affiliation{Center for Cyber-Physical Systems (C2PS), Khalifa University, 127788 Abu Dhabi, UAE.}
\author{M. Asjad}
\email{asjad\_qau@yahoo.com}
\affiliation{Department of Mathematics, Khalifa University, Abu Dhabi 127788, United Arab Emirates.}
\begin{abstract}
We propose a scheme to achieve magnon (photon) blockade by using magnon squeezing within a cavity magnomechanical system under weak pump driving.  Under ideal conditions, we observe a substantial magnon blockade effect, as well as simultaneous photon blockade. Moreover, both numerical and analytical results match perfectly, providing robust evidence of consistency. In addition to calculating optimal parametric gain and detuning values, we can improve the second-order correlation function. The proposed scheme will be a pioneering approach towards magnon (photon) blockade in experimental cavity magnomechanical systems.
\end{abstract} 
\maketitle 
\section{Introduction}
Photon statistics are integral to quantum optics \cite{RJGlauber2006}. Initially unveiled by Hanbury Brown \cite{RHBrown1952, RHBrown1956} and confirmed by Kimble et al. \cite{HJKimble1977}, photon antibunching illustrates light's intriguing properties. An enthralling phenomenon, photon blockade (PB), arises within nonlinear media, where the presence of a single photon obstructs the entry of another photon \cite{AImamoglu1997}.  PB's theoretical relevance extends to diverse nonlinear optical systems, encompassing Kerr-type nonlinear cavities \cite{JQLiao2010} and optomechanical setups \cite{PRabl2011, aa1, aa2, JQLiao872013, JQLiao882013, HWang2015, GLZhu2018, FZou2019}. Experimental manifestations of PB materialize in contexts like an optical cavity coupled to a confined atom \cite{KMBirnbaum2005} and a photonic crystal cavity linked with a quantum dot \cite{AFaraon2008, AReinhard2012, KMuller2015}. The ability to generate single-photon sources is of paramount importance for quantum information technologies \cite{EKnill2001,PKok2007}. The photon blockade effect plays a crucial role in achieving this objective as it produces strongly anti-bunched photons. This effect enables the generation of sub-Poissonian light even when the system is driven by a classical light field.

In recent years, the photon blockade effect has gained significant importance in quantum information processing and quantum communication. Currently, there are two well-known mechanisms for achieving a strong photon blockade effect: unconventional photon blockade (UCPB) \cite{TCHLiew2010, MBamba2011, MBajcsy2013}, which relies on destructive quantum interference between two different quantum transition paths from the ground state to a two-excited state, and conventional photon blockade (CPB) \cite{AImamoglu1997, WLeonski1994, LTian1992} which relies on large nonlinearities to alter the energy-level structure of the system. Experimental implementations of both UCPB and CPB mechanisms have been reported in \cite{KMBirnbaum2005, AFaraon2008, AJHoffman2011, HJSnijders2018, CVaneph2018}. 

Cavity magno-mechanics has emerged as a promising research area, providing a solid platform for studying ferrimagnetic crystals such as YIG spheres coupled with microwave cavities \cite{DDLachanceQuirion2019, aa3, HYYuanArxiv}. This coupling is achieved through magnetostrictive force, which combines the ferromagnet's vibratory deformation mode with the microwave cavity mode, and the magnetic dipole interaction. When the mechanical mode frequency is significantly lower than the magnon frequency, a dispersive interaction known as the magnetostrictive interaction occurs, which can be compared to radiation pressure in larger ferromagnetic systems \cite{XZhang2016,Asjad20233, ZYFan2022}. Such interactions enable the coupling of mechanical and magnetic degrees of freedom in nanoscale devices, facilitating the manipulation and control of magnetic properties through mechanical means. By exploiting this coupling, it is possible to control and manipulate the flow of magnons within the system. Magnon blockade refers to the phenomenon where one magnon inhibits additional magnons through a given channel. 

In this paper, we present a scheme to investigate the magnon (photon) blockade effect in an opto-magnomechanical system with magnon squeezing under the weak driving limit, focusing on the second-order correlation function of an opto-magnomechanical system. Analytically and numerically, we demonstrate that a magnon (photon) blockade can be generated by controlling the nonlinear parameter value and selecting appropriate system parameters to generate a magnon (photon) blockade. Furthermore, we investigate the robustness of the magnon blockade in the presence of pure dephasing in the system.

This paper is organized as follows: in Section \ref{sec2}, we provide an overview of the opto-magnomechanical system with magnon squeezing and present its corresponding Hamiltonian. In Section \ref{sec3}, we investigate the magnon (photon) blockade using the second-order correlation function. We analyze the blockade effect under optimal parameters using analytical and numerical methods. Additionally, we explore system dynamics and discuss the obtained results. Finally, we present our conclusions in Section \ref{sec4}.

\section{The Model Hamiltonian} \label{sec2}
The microwave cavity and the YIG sphere, which include the magnon-photon and magnon-phonon-induced interactions of the magnetic dipole and magnetostriction, respectively, make up the magnomechanical system of the cavity, as depicted in Figure 1. The YIG sphere's geometry deforms due to the magnetic change brought on by the magnon's excitation. This creates vibratory modes (phonons) in the sphere, and vice versa. In the YIG sphere position, the external drive magnetic field is situated along the y axis, the microwave cavity's additional polarization magnetic field is positioned along the z axis, and the cavity mode's magnetic field is oriented along the x axis. The magnetostrictive effect causes the photon and magnon to couple. One can change the magneton-photon coupling force by using an extra magnetic field and moving the YIG sphere. Due to the high spin density $(4.2\times 10^{21}cm^{-3})$ of the magnon in the YIG sphere, strong coupling to the microwave cavity is possible. The total Hamiltonian describing the system in the rotating frame approximation is given by ($\hbar=1$)
\begin{eqnarray}
\mathcal{H} &=& \Delta_c c^{\dag}c + \Delta_m m^{\dag}m + \omega_b b^{\dag}b - g_{mb}(b + b^{\dag})m^{\dag}m \nonumber \\
&+& g_{mc}(m^{\dag} c + c^{\dag} m)+ \ic\lambda \big((m^{\dagger})^2\e^{\ic\theta} - m^2\e^{-\ic\theta}\big) \nonumber \\
&+& \mathcal{E}(m^{\dag} + m),\label{eq:1a} 
\end{eqnarray} 
\begin{figure}[!htb]
\centering
\includegraphics[width=0.45\textwidth, height=1.5 in]{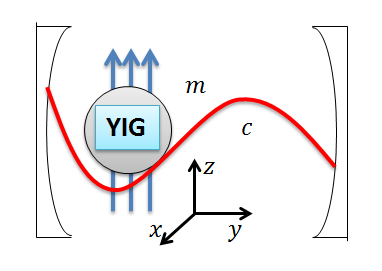}
\caption{A single-mode cavity contain inside a $YIG$ sphere with magnon squeezing. In order to improve the magnomechanical coupling, the magnon mode is directly driven by a microwave source (not shown), which causes a large number of spins in a macroscopic ferrimagnet to move together in a coordinated motion. Another force that drives the cavity is an electromagnetic field with amplitude $\mathcal{E}$. Magnons and cavity photons are coupled through magnetic dipole interaction, while phonons and magnons are connected through magnetostrictive (radiation pressure-like) interaction.}
\label{schema}
\end{figure}
where $\Delta_c = \omega_c - \omega_l$ and $\Delta_m = \omega_m - \omega_l$ are the detunings of the microwave cavity mode and magnon mode, respectively. When a unitary transformation $\mathcal{U} =\exp[g_{mb} (b - b^{\dag}) / \omega_b]$ is applied to $\mathcal{H}$ and in the case of $g_{mb} \ll \omega_b$, The unitary transformation $\mathcal{U}$ decouples the two mechanical and the optical modes in the special case of \textit{weak optomechanical coupling}, i.e. $g_{mb}/\omega_b\ll 1$  thus we can safely ignore the mechanical mode. Therefore, the Hamiltonian can be further expressed as
\begin{eqnarray} \label{eq:2} 
\mathcal{H}_1 &=& \Delta_c c^{\dag}c + \Delta_m m^{\dag}m + g_{mc}(m^{\dag} c + c^{\dag} m) - \mu (m^{\dag}m)^2 \nonumber \\
&+& \ic \lambda \big((m^{\dagger})^2\e^{\ic\theta} - m^2\e^{-\ic\theta}\big) + \mathcal{E}(m^{\dag} + m),\label{eq:1} 
\end{eqnarray} 
where $\mu=g_{mb}^2/\omega_b$ is the Kerr-type nonlinear strength. the Hamiltonian of non-Hermitian system is given by
\begin{equation} \label{eq:5} 
\mathcal{H}_2=\mathcal{H}_1-\ic\frac{\kappa_c}{2}c^{\dagger}c-\ic\frac{\kappa_m}{2}m^{\dagger}m.
\end{equation}
\section{Magnon (photon) blockade} \label{sec3}
We concentrate on the analytical solution of the non-Hermitian Schr\"odinger equation as well as numerical solution using Master equation. In
 addition, we talk about the weak coupling regime's ability to realize a strong magnon (photon) blockade ($g_{mb}\ll \omega_b$). The analytical expression of the correlation function can be obtained by solving the Schr\"odinger equation in the weak driving limit, i.e. $ \ic\frac{\partial |\varphi (t)\rangle}{\partial t}=\mathcal{H}_2|\varphi (t)\rangle$, where $|\varphi (t)\rangle$ denotes the system's current state. 

The low-excitation subspace has a limit on the evolution space of up to 2, i.e., $\mathcal{D}=\{|q,r\rangle/q+r\leq 2 \}$. The state of the system with the bare-state bases are : $|\varphi (t)\rangle=\sum_{q, r}^{q+r \leq 2}\mathcal{P}_{qr} (t) |q, r\rangle$, where $\mathcal{C}_{qr} (t)$ is the amplitude of the bare state's probability. The symbol $|q, r\rangle$ with $q$ denotes the presence of magnons in the cavity, while $r$ denotes the presence of photons in the cavity. The Schr\"odinger equation generates the subsequent set of linear differential equations for the probability amplitudes $\mathcal{P}_{qr} (t)$
\begin{eqnarray} \label{eq:7} 
\ic\frac{\partial \mathcal{P}_{00}}{\partial t} &=& \mathcal{E}\mathcal{P}_{10}-\ic\sqrt{2}\lambda\mathcal{P}_{20},\nonumber\\
\ic\frac{\partial \mathcal{P}_{10}}{\partial t} &=& \mathcal{E}\mathcal{P}_{00}+(\Delta'_m - \mu)\mathcal{P}_{10}+g_{mc}\mathcal{P}_{01}+\sqrt{2}\mathcal{E}\mathcal{P}_{20},\nonumber\\
\ic\frac{\partial \mathcal{P}_{01}}{\partial t} &=& g_{mc}\mathcal{P}_{10}+\Delta'_c\mathcal{P}_{01}+\mathcal{E} \mathcal{P}_{11},\nonumber\\
\ic\frac{\partial \mathcal{P}_{11}}{\partial t}&=& \sqrt{2}g_{mc}\mathcal{P}_{20}+\mathcal{E}\mathcal{P}_{01}+(\Delta'_c + \Delta'_m - \mu)\mathcal{P}_{11}\nonumber\\
&+&\sqrt{2}g_{mc}\mathcal{P}_{02},\nonumber\\
\ic\frac{\partial \mathcal{P}_{02}}{\partial t}&=& \sqrt{2}g_{mc}\mathcal{P}_{11}+2\Delta'_c\mathcal{P}_{02},\nonumber\\
\ic\frac{\partial \mathcal{P}_{20}}{\partial t}&=& \sqrt{2}\mathcal{E}\mathcal{P}_{10}+2(\Delta'_m - 2\mu)\mathcal{P}_{20}+\sqrt{2}g_{mc}\mathcal{P}_{11}\nonumber\\
&+& \ic\lambda\sqrt{2} \e^{i\theta}\mathcal{P}_{00},
\end{eqnarray} 
where $\Delta'_c=\Delta_c-\ic\kappa_c/2$ and $\Delta'_m=\Delta_m-\ic\kappa_m/2$. The above set of equations (Eq. \ref{eq:7}) can be solved analytically in the steady-state by setting $\partial \mathcal{P}_{qr}/\partial t$ to zero. Under the condition of weak driving ($\mathcal{E}\ll \kappa$), the probability amplitudes satisfy the relation $|\mathcal{P}_{00}|\simeq 1\gg |\mathcal{P}_{01}|,|\mathcal{P}_{10}|\gg |\mathcal{P}_{11}|,|\mathcal{P}_{02}|,|\mathcal{P}_{20}|$.  The corresponding steady-state values of the probability amplitudes, assuming $\Delta_c=\Delta_m=\Delta$, $\kappa_c=\kappa_m=\kappa$, and $\theta=0$, are given by $\mathcal{P}_{10}=\mathcal{E}\Delta'/\chi$, $\mathcal{P}_{01}=g_{mc}\mathcal{E}/\chi$,   $\mathcal{P}_{20}=-\mathcal{A}/\mathcal{B}$ and $\mathcal{P}_{02}=-\mathcal{C}/\mathcal{B}$, where $\chi = g^2_{mc}+\mu\Delta' -\Delta'^2 $,
\begin{eqnarray}
\mathcal{A}&=& \mathcal{E}^2 \Delta'^2 (2\Delta'-\mu) -i \lambda \chi \left(\chi-\Delta'^2\right) \nonumber\\
\mathcal{B}&=&\sqrt{2} \chi \left[\Delta' (\Delta'-2\mu) (2\Delta'-\mu) -2 g^2_{mc} (\Delta'-\mu)   \right]\nonumber\\
\mathcal{C}&=&g^2_{mc} [2\mathcal{E}^2 (\Delta'-\mu) +i\lambda \chi].
\end{eqnarray}
with $\Delta'_m=\Delta'_c=\Delta'$. By examining the magnon (photon) statistics, we can investigate the occurrence of magnon (photon) blockade. The second-order correlation function in the steady state provides valuable information in this regard. Specifically, we analyze $g_m^{(2)}(0)$ and $g_c^{(2)}(0)$, which are respectively given by:
\begin{equation} \label{eq:18}
g_m^{(2)}(0)=\frac{2|\mathcal{P}_{20}|^2}{|\mathcal{P}_{10}|^4}\quad;\quad g_c^{(2)}(0)=\frac{2|\mathcal{P}_{02}|^2}{|\mathcal{P}_{01}|^4}.
\end{equation}
The condition $g_j^{(2)}(0)>1$ ($j=m, c$) corresponds to the magnon (photon) bunching effect, indicating a higher probability of detecting multiple magnons (photons) simultaneously. Conversely, when $g_j^{(2)}(0)<1$ ($j=m, c$), it signifies magnon (photon) antibunching, where there is a reduced likelihood of detecting multiple magnons (photons) together. Moreover, the magnon blockade effect occurs in the cavity when $g_m^{(2)}(0)=0$, indicating the complete suppression of magnon emission. Similarly, the strong photon blockade effect can be observed in the cavity when $g_c^{(2)}(0)=0$, indicating the complete suppression of photon emission. The Master equation approach, which enables us to perform numerical calculations, can be expressed as follows: 
\begin{equation} \label{eq:21a} 
\frac{\dd \rho}{\dd t}=-\ic [\mathcal{H}_1,\rho]+\mathcal{L}_{c}(\rho)+\mathcal{L}_{m}(\rho),
\end{equation}
where $\mathcal{L}_{c}(\rho)=\frac{\kappa}{2}(2c\rho c^+-c^{\dagger}c\rho -\rho c^{\dagger} c)$ and $\mathcal{L}_{m}(\rho)=\frac{\kappa}{2}(2m\rho m^+-m^{\dagger}m\rho -\rho m^{\dagger} m)$. $\mathcal{H}_1$ is the Hamiltonian given by Eq.(\ref{eq:2}). $\kappa$ is the corresponding decay rate of the mode.
\begin{figure}[!htb]
\begin{center}
\includegraphics[width=.43\textwidth]{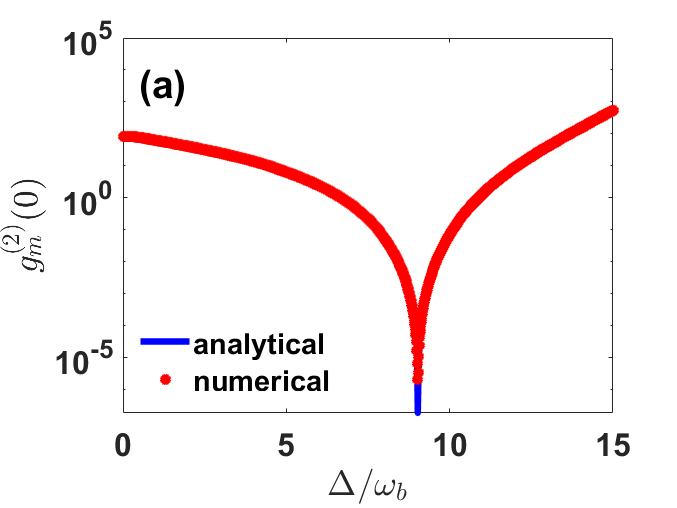}
\includegraphics[width=.43\textwidth]{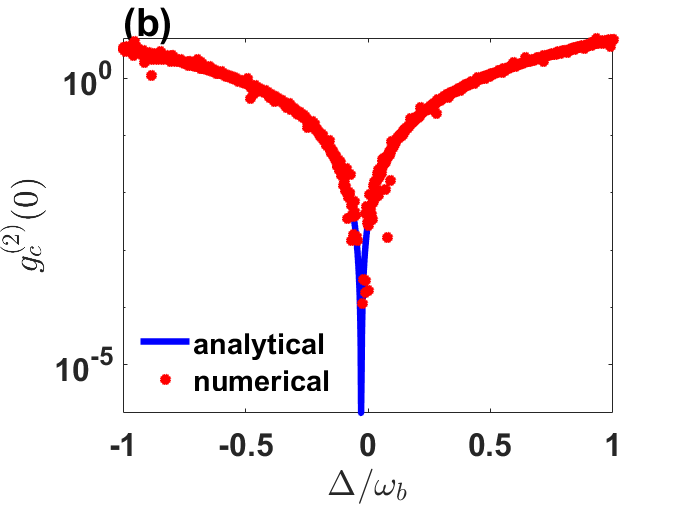}
\end{center}
\caption{Plot of the equal-time second-order correlation function $g_j^{(2)}(0)$ $(j=m,c)$ versus the detuning $\Delta$ with $\lambda=\lambda^{opt}$. The blue solid line represents the analytical results, and the red dotted line represents the numerical results, respectively.  See the text for value of other parameters.}
\label{NTh}  
\end{figure}
\subsection{RESULTS AND DISCUSSIONS} \label{sec4}
The achieved magnon and photon blockade effects in this study correspond to the Unconventional Cross-Phase-Modulation Blockade (UCPB) mechanism, which arises from destructive quantum interference. The following parameter values were used: $\kappa/2\pi = 1\times 10^6$ Hz, $\omega_b = \kappa$, $\mathcal{E}=0.01\kappa$, $g_{mb} = 3\kappa$, and $g_{mc} = 0.5\kappa$. The optimal parameter pairs $\lambda$ and $\Delta$ can be obtained alongside the other fixed parameters. The real solutions for the optimal parameter values are as follows: (i) for magnon blockade, ${ \Delta^{opt} \to 9.03 \omega_b , \lambda^{opt} \to 2\times 10^{-4}\omega_b }$, and (ii) for photon blockade, ${ \Delta^{opt} \to -0.03 \omega_b , \lambda^{opt} \to 4\times 10^{-4}\omega_b }$.
\begin{figure}[!htb]
\begin{center}
\includegraphics[width=.43\textwidth]{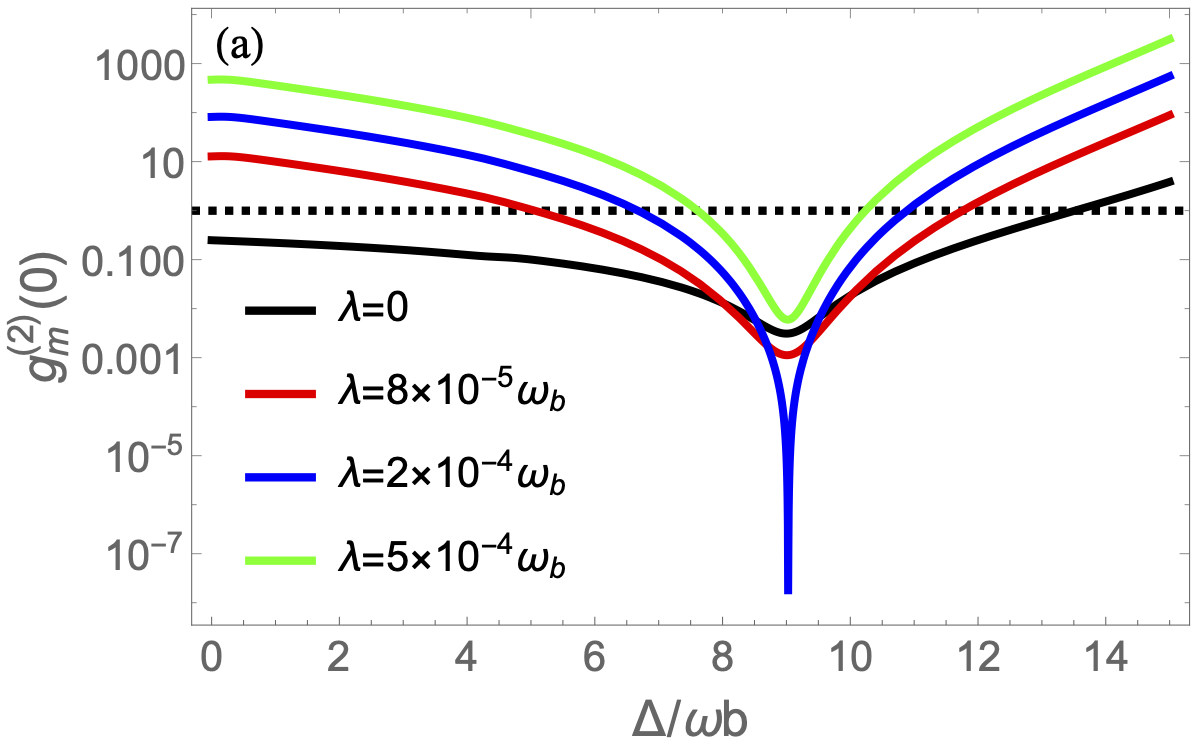}
\includegraphics[width=.43\textwidth]{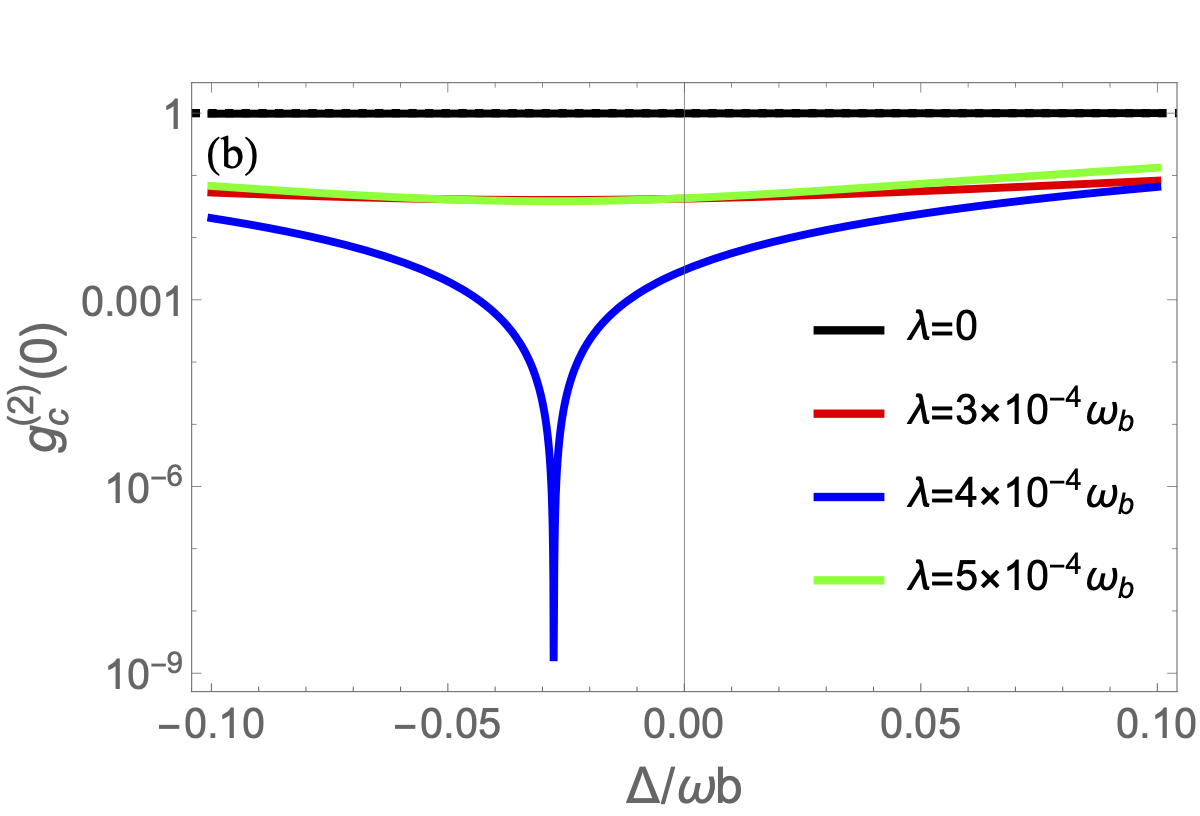}
\end{center}
\caption{Plot of the analytical results of the equal-time second-order correlation function $g_j^{(2)}(0)$ $(j=m,c)$ versus the detuning $\Delta$ for different values of the parameter $\lambda$. The dashed horizontal line delimits the region under which represents the photon-magnon antibunching in the cavity. See the text for value of other parameters.}
\label{mB}
\end{figure}
In Fig. \ref{NTh}, we present the equal-time second-order correlation function $g_m^{(2)}(0)$ and $g_c^{(2)}(0)$ as a function of the normalized detuning $\Delta/\omega_b$ for optimal values obtained from both analytic and numerical solutions. In Fig. \ref{NTh}(a), the second-order correlation function exhibits a strong unconventional magnon blockade at the precise value of $\Delta = 9.03 \omega_b$ for $\lambda = 2\times 10^{-4}\omega_b$, as accurately predicted by the analytically calculated optimal parameters. Similarly, in Fig. \ref{NTh}(b), unconventional photon blockade is observed at $\Delta = -0.03 \omega_b$ for $\lambda = 4\times 10^{-4}\omega_b$. This behavior can be attributed to destructive quantum interference between different excitation paths. Furthermore, in Fig. \ref{NTh}(a), another small dip is observed with $g_m^{(2)}(0)>1$, indicating a bunching effect. The numerical results are in excellent agreement with the analytical results, verifying their consistency under weak driving conditions. It is noteworthy that the optimal value leading to a complete photon-magnon blockade is achieved through analytical calculations.

Figures \ref{mB}(a)-\ref{mB}(b) display the second-order correlation function $g_j^{(2)}(0)$ ($j=m,c$) as a function of the detuning $\Delta$ for different values of $\lambda$, while keeping other parameters fixed. It is observed that the values of $g_m^{(2)}(0)$ are significantly smaller for the listed ${\Delta, \lambda}$ pairs than for others. Specifically, Figure \ref{mB}(a) demonstrates a strong unconventional magnon blockade effect at $\Delta = 9.03 \omega_b$ with $\lambda = 2\times 10^{-4} \omega_b$, while Figure \ref{mB}(b) shows a strong unconventional photon blockade effect at $\Delta = -0.03 \omega_b$ with $\lambda = 4\times 10^{-4} \omega_b$. These results agree with the analytically calculated optimal parameters. We note that when $\lambda = 0$, implying the absence of magnon squeezing, the unconventional magnon blockade (UMB) cannot be achieved, whereas the conventional magnon blockade (CMB) is attainable at $\Delta \approx 4.5 \omega_b$, as shown in Figure \ref{mB}(a). Additionally, for instance, when $\lambda=500$ Hz, a small dip appears corresponding to the conventional magnon blockade, as depicted in Figure \ref{mB}(a). On the other hand, in Figure \ref{NTh}(a), when $\lambda = \lambda^{opt}$, another small dip exists with $g_m^{(2)}(0)>1$, indicating a bunching effect. As anticipated in Figure \ref{NTh}(b), the different curves lack a sharp dip, implying photon blockade cannot be achieved. 
\begin{figure}[!htb]
\begin{center}
\includegraphics[width=.43\textwidth]{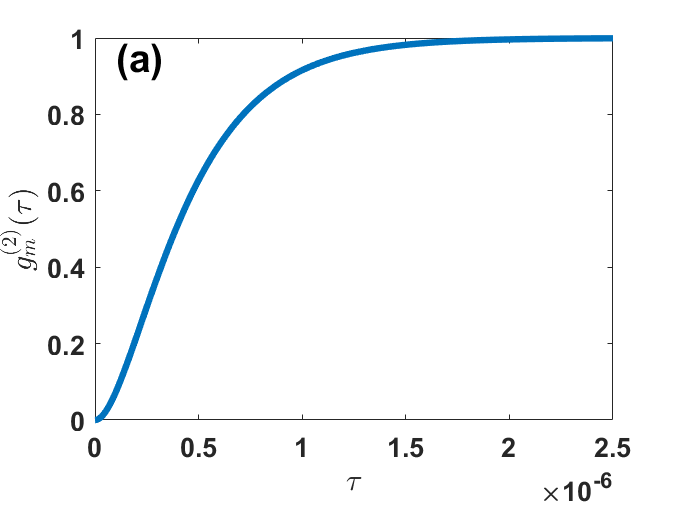}
\includegraphics[width=.43\textwidth]{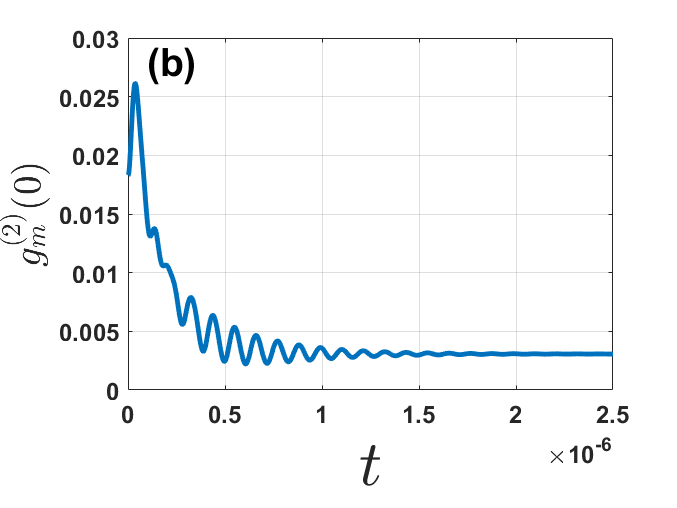}
\end{center}
\caption{(a) Time evolution of the second-order correlation function, $g_m^{(2)}(\tau)$ $(t = 0)$, (b) Plot of $g_m^{(2)}(0)$ as a function of $t$ $(\tau = 0)$. See the text for value of other parameters.}
\label{phabB}
\end{figure}
\begin{figure}[h!]
\begin{center}
\includegraphics[width=.44\textwidth]{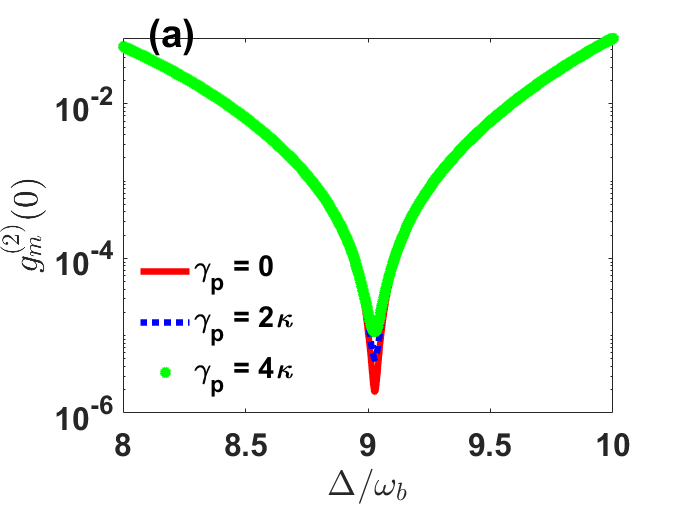}
\includegraphics[width=.44\textwidth]{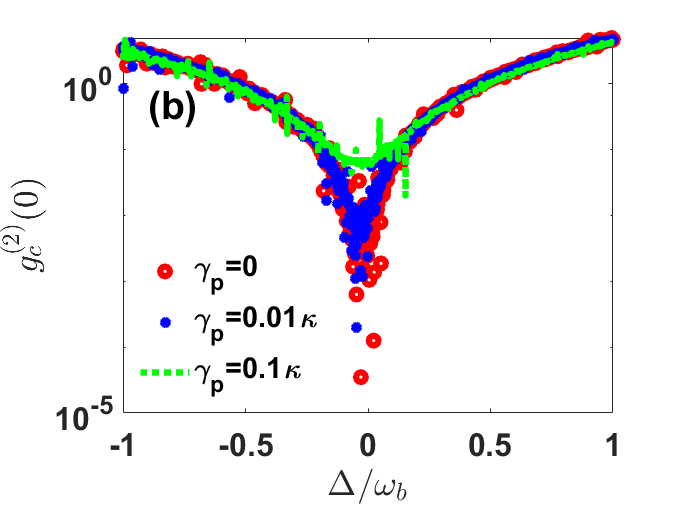}
\end{center}
\caption{Plot of the equal-time second-order correlation function $g_j^{(2)}(0)$ $(j=m,c)$ versus the detuning $\Delta$ for different pure-dephasing rates $\gamma_p$. See the text for value of other parameters.}
\label{dfz}
\end{figure}

In Fig. \ref{phabB}(a), we depict the time evolution of the second-order correlation function $g_m^{(2)}(\tau)$, which is calculated using the expression:
\begin{equation}
g_m^{(2)}(\tau)=\frac{\langle m^\dag (t) m^\dag(t+\tau)m(t+\tau)m (t) \rangle}{\langle m^\dag (t)m (t) \rangle^2}.
\end{equation}
This quantity is a significant measure of the photon statistical properties. It is proportional to the joint probability of detecting one photon at $t = 0$ and detecting the next photon at $t = \tau$. It can be observed that for delay times $\tau > 0$, $g_m^{(2)}(\tau)$ is greater than $g_m^{(2)}(0)$. This indicates that the emitted photons exhibit blockade behavior and possess sub-Poissonian statistics. Notably, as the delay time $\tau$ increases, $g_m^{(2)}(\tau)$ also increases and approaches its maximum value (unity) at $\tau = 1.6\mu$s, which corresponds approximately to the lifetime of the photons in the cavity. This signifies that with increasing time delay, photons become more coherent and exhibit stronger correlation.

In Fig. \ref{phabB}(b), we present the time evolution of the second-order correlation function, $g_m^{(2)}(0)$. It can be observed that $g_m^{(2)}(0)$ decreases as time increases. Additionally, there is a significant oscillation in the initial short time, which is a result of the classical driving field. However, this oscillation gradually stabilizes, reaching a steady state. Throughout the time evolution, the second-order correlation function remains below 1 ($g_m^{(2)}(0) \ll 1$), ensuring the validity of a strong magnon blockade.
\subsection{Pure dephasing effects} 
Pure dephasing can have a negative impact on the magnon (photon) blockade characteristics and introduce unwanted decoherence in the system. Hence, we investigate the influence of pure dephasing on cavity photon antibunching characteristics. To model the effects of pure dephasing, an additional Lindblad term of the form $\mathcal{L}_p(\rho) = \frac{\gamma_p}{2}[2c^{\dag} c\rho c^\dag c - (c^{\dag} c)^2 \rho - \rho(c^{\dag} c)^2]$ is included in the master equation, where $\gamma_p$ represents the pure dephasing rate of the cavity mode.

In Fig. \ref{dfz}(a)-(b), we present the plots of the equal-time second-order correlation function $g_j^{(2)}(0)$ $(j=m,c)$ against the detuning $\Delta$ for various values of the pure-dephasing rates $\gamma_p$. We observe that even after including pure-dephasing induced losses in the system, photon blockade and magnon blockade remain robust. When $\gamma_p = 0$, $g_m^{(2)}(0)$ exhibits strong sub-Poissonian quantum statistics at the detuning $\Delta = \omega_b$. Furthermore, $g_m^{(2)}(0)$ increases with increasing decay rates $\gamma_p$. In other words, for higher pure-dephasing rates, $g_m^{(2)}(0)$ approaches classical Poissonian statistics similar to a thermal source, as expected (Fig. \ref{dfz}(a)). Similar features can be observed for $g_c^{(2)}(0)$ for this value of $\gamma_p$ (Fig. \ref{dfz}(b)).
\section{Conclusion}\label{sec5}
In conclusion, we have proposed a scheme to achieve strong magnon (photon) blockade in an opto-magnomechanical system under weak pump driving with magnon squeezing. We investigated the magnon (photon) correlations in terms of the second-order correlation function. The magnon (photon) blockade is realized under optimal parameters, and we have found complete agreement between the analytical and numerical results. Additionally, we have analyzed that the magnon (photon) blockade effect can be achieved by adjusting the nonlinear parameter and selecting appropriate parameters. We discussed both unconventional and conventional magnon blockade. Furthermore, we have studied the time-delay second-order correlation function in the presence of magnon squeezing. Finally, we have shown that the magnon blockade remains robust even in the presence of pure-dephasing induced losses in the system.

\end{document}